\documentclass{iopart}

\usepackage{setstack}
\usepackage{iopams}
\usepackage{graphicx}
\usepackage[dvips]{color}

\begin{document}

\title[Ramsauer approach to Mie scattering of light on spherical particles] 
{Ramsauer approach to Mie scattering of light on spherical particles}
\author{K Louedec, S Dagoret-Campagne, and M Urban}
\address{LAL, Univ Paris-Sud, CNRS/IN2P3, Orsay, France.}
\ead{louedec@lal.in2p3.fr}

\begin{abstract}
The scattering of an electromagnetic plane wave by a spherical particle was solved analytically by Gustav Mie in 1908. The Mie solution is expressed as a series with very many terms thus obscuring the physical interpretations of the results. The purpose of the paper is to try to illustrate this phenomenon within the Ramsauer framework used in atomic and nuclear physics. We show that although the approximations are numerous, the Ramsauer analytical formulae describe fairly well the differential and the total cross sections. This allows us to propose an explanation for the origin of the different structures in the total cross section.
\end{abstract}

\submitto{\PS}
\maketitle

\section{Introduction}
\label{sec:introduction}
The subject of light scattering by small particles is present in several scientific areas such as astronomy, meteorology or biology~\cite{Hulst}. The first model by Lord Rayleigh in 1871 dealt with light scattering by particles whose dimensions are small compared to the wavelength. A significant improvement came with the Mie solution~\cite{Mie} which describes light scattering by spherical particles of any size. This extension is important in astronomy and meteorology where light can go through aerosols which are particles dispersed in the atmosphere. In Section~\ref{sec:mie}, we study the Mie series for the light scattering on dielectric spheres in air. In an apparently different domain, the scattering of a low energy electron by atoms studied by Ramsauer~\cite{Ramsauer} showed surprising structures, and it took several years before the solution was imagined by Bohr: describe the electron as a plane wave~\cite{e-atome1,e-atome2,Karwasz}! The Ramsauer effect is certainly the first phenomenon showing the wave properties of matter. This framework has since then been used to describe very different types of collisions such as atom-atom~\cite{atome-atome1} or even neutron-nucleus~\cite{neutron-noyau1,neutron-noyau2,neutron-noyau3,Peterson,Abfalterer}. Our idea is that since light behaves like a wave, it should be also possible to apply Ramsauer's ideas to the light scattering by dielectric droplets. The Ramsauer effect is described in Section~\ref{sec:ramsauer}.
Finally, in Section~\ref{sec:application}, we compare the predictions of Mie and Ramsauer for the total scattering cross section of light over a rain drop.

\section{The Mie predictions for light over a sphere of non absorbing dielectric}
\label{sec:mie}
The Mie solution is detailed in the \ref{app:mie}. Let $R$ be the radius of the sphere, let $n$ be the index of refraction of the dielectric, and let $\sigma_{\rm tot}$ be the total cross section. In the context of light scattering, the extinction efficiency factor $Q_{\rm e} = \sigma_{\rm tot}/(\pi R^2) $ is often used.

\begin{figure}
\centering
\begin{tabular}{c}
\includegraphics[width = 0.75\textwidth]{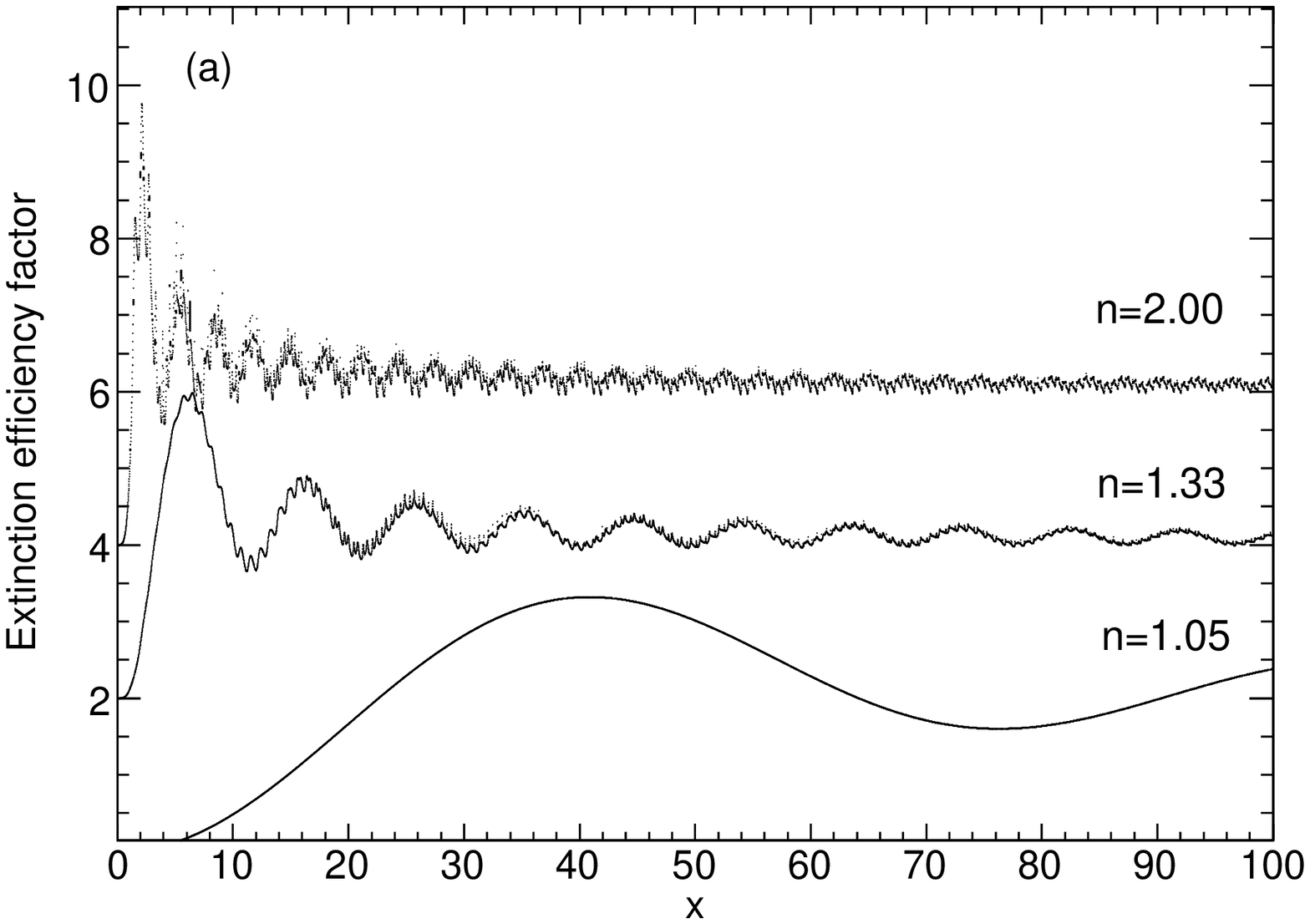} \\
\includegraphics[width = 0.75\textwidth]{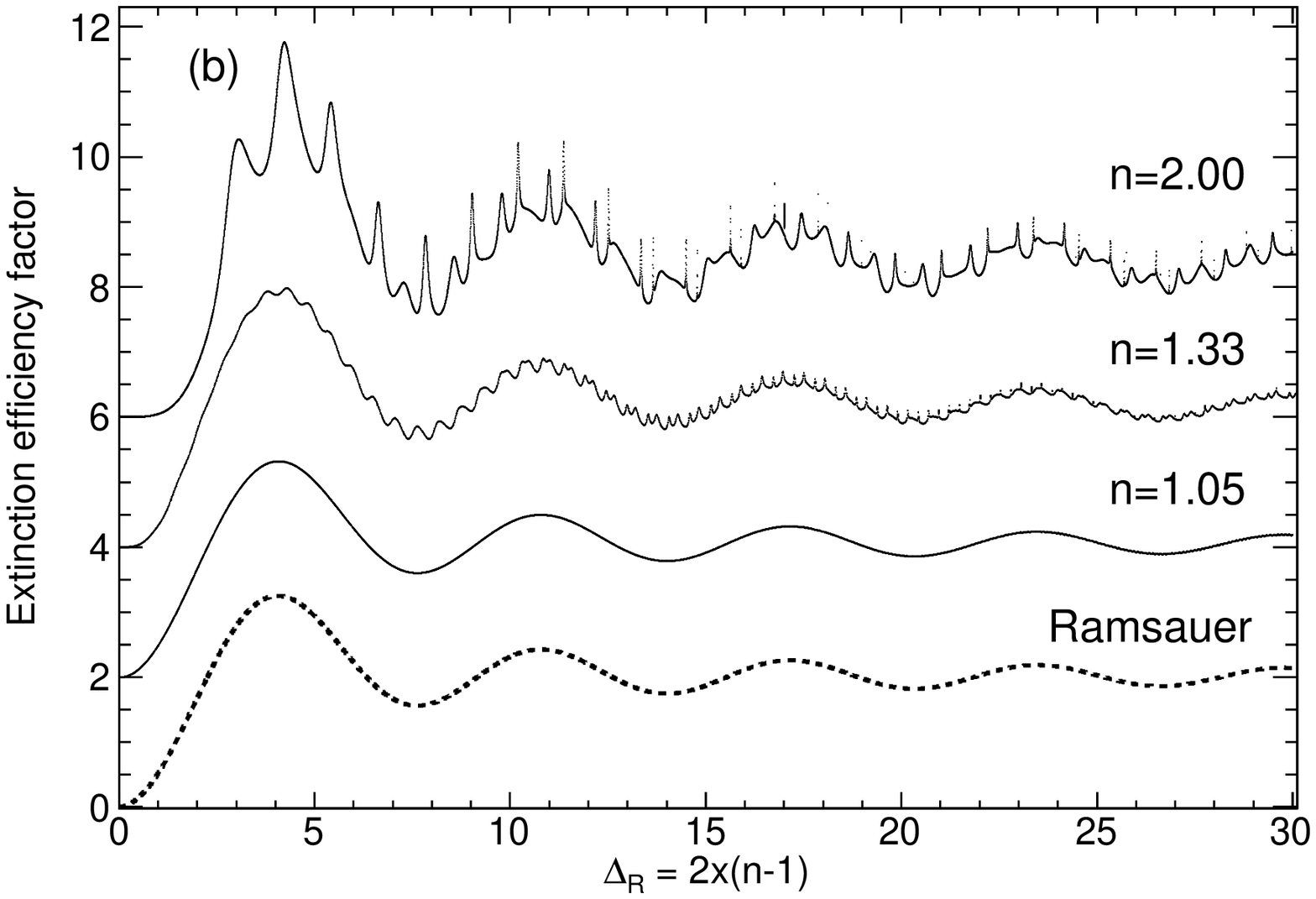}
\end{tabular}
\caption{The extinction efficiency factor versus (a) $x$ and (b) $\Delta_{\rm R}=2x\,(n-1)$, for non absorbing spherical particles with relative refractive indices $n=1.05,\: n=1.33$, and $n=2.00$. $x$ is given by the relation $x = 2 \pi R/ \lambda = k_{\rm out}R$. Ramsauer solution for $n=1.05$ is also given in low panel. In (a) and (b), the vertical scale applies only to the lowest curve, the others being successively shifted upward by 2.}
\label{fig:1}
\end{figure}

Figure~\ref{fig:1}.(a) shows, for three values of the index of refraction, $Q_{\rm e}$ as a function of the size parameter $x = 2 \pi R/ \lambda = k_{\rm out}R$, where $k_{\rm out}$ is the wave number of light outside the sphere. Each curve is characterized by a succession of maxima and of minima with superimposed ripples. The amplitude of the large oscillations and of the ripples grows with $n$. In the next sections we will show that all these features can be understood with the Ramsauer approach.

\section{Description of the Ramsauer effect}
\label{sec:ramsauer}

\begin{figure}[ht]
\centering
\includegraphics[width = 0.75\textwidth]{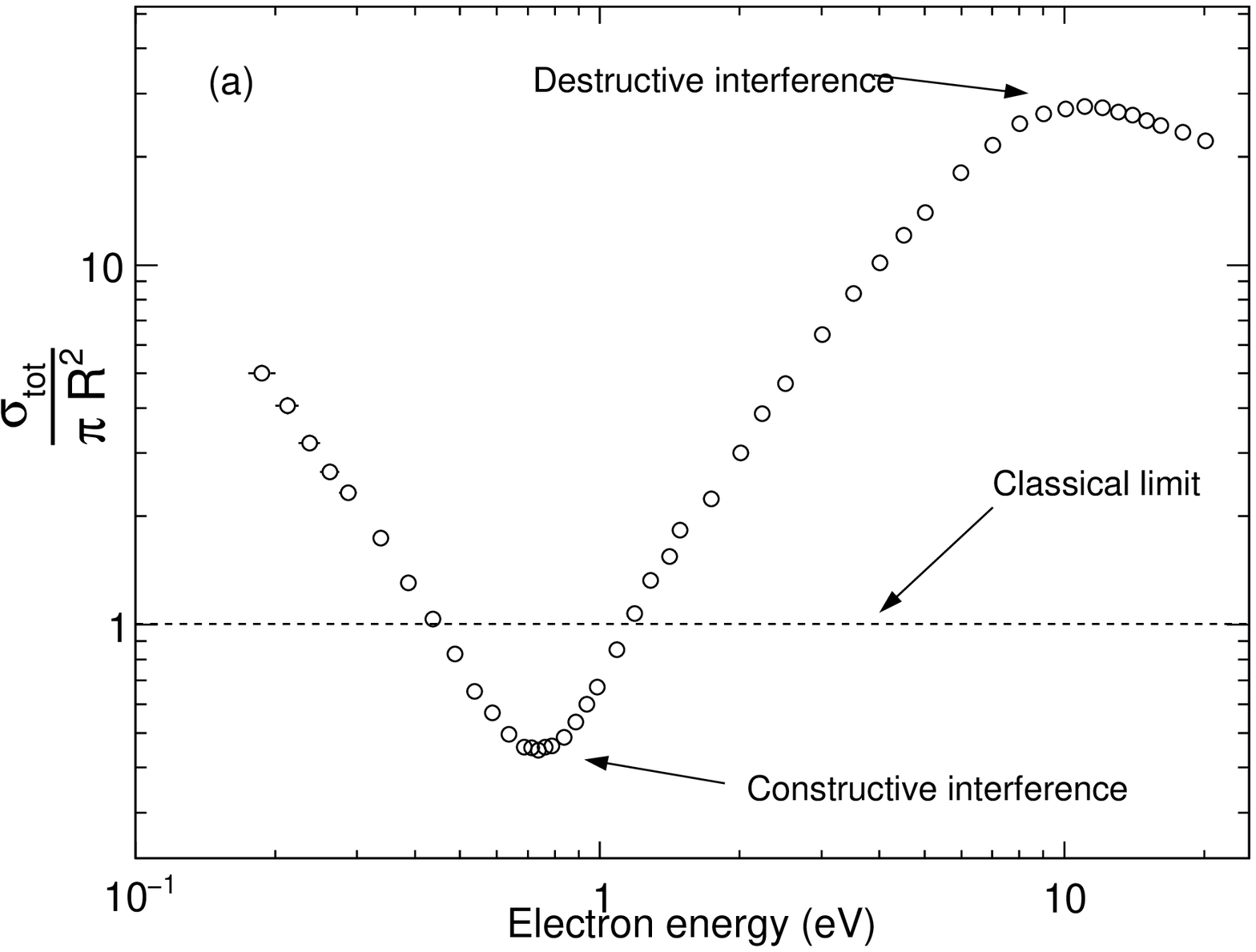}
\includegraphics[width = 0.75\textwidth]{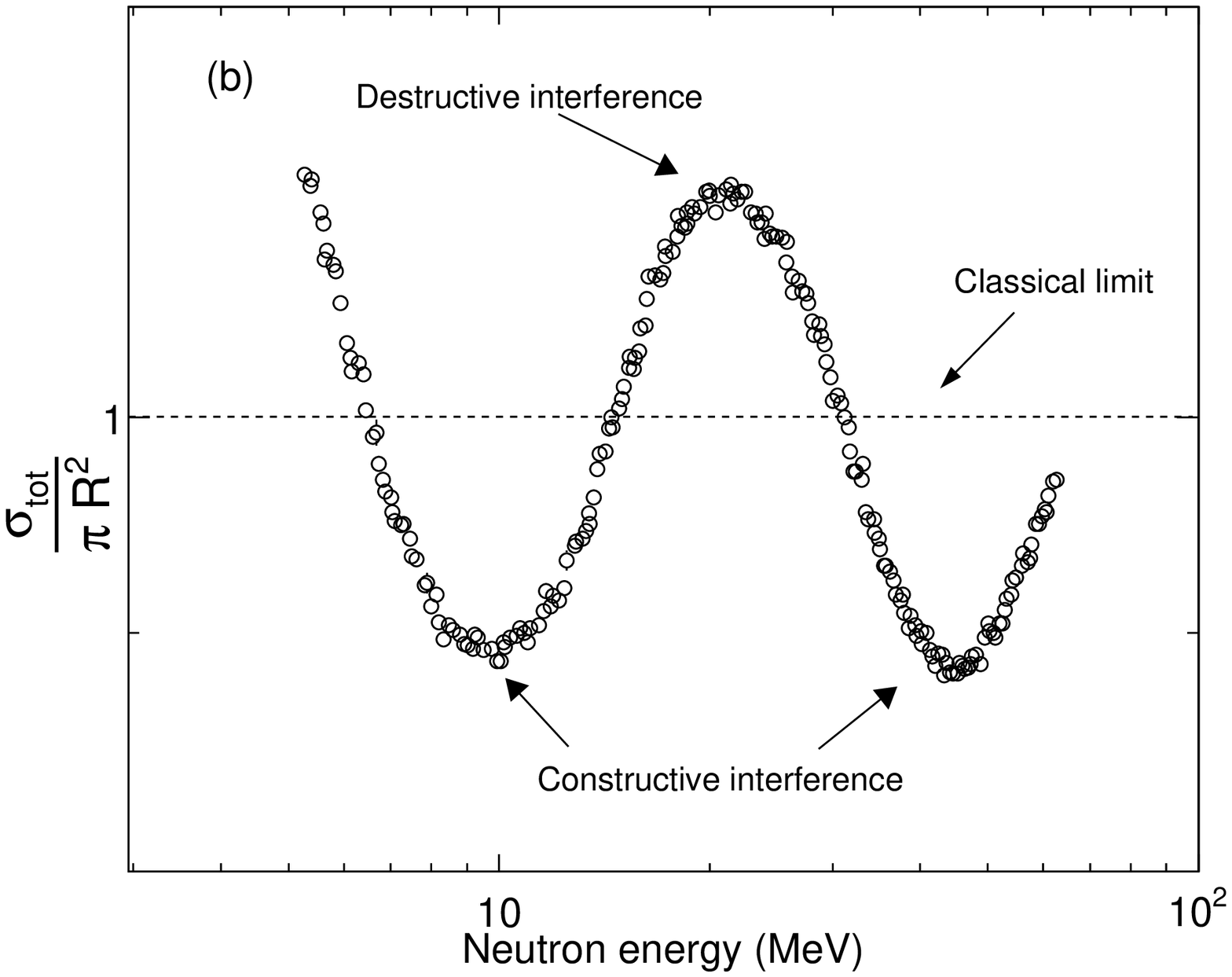}
\caption{(a)~Electron-Krypton normalized total cross section versus energy \cite{Karwasz}. (b)~Normalized Neutron-Lead total scattering cross section versus energy \cite{Abfalterer}.}
\label{fig:2}
\end{figure}

The Ramsauer effect was discovered in 1921~\cite{Ramsauer} while studying electron scattering over Argon atoms. The total cross section versus the electron energy showed a surprising dip around $1$\,eV. In Figure~\ref{fig:2} we see recent measurements of electron over Krypton and neutron over Lead nucleus. 

\begin{figure}
\centering
\includegraphics{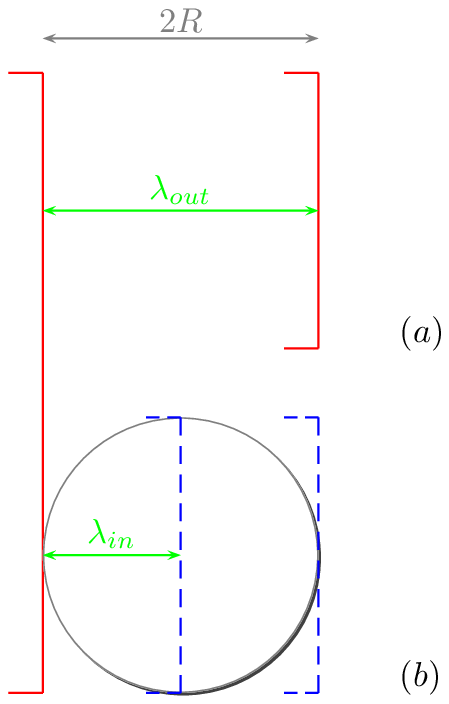}
\caption{Qualitative picture of the Ramsauer phenomenon. The wavelength of the light is supposed to be reduced in the dielectric. This picture shows the case where the contraction of the wavelength between (a) outside and (b) inside the medium (dashed line) is such that they come out in phase. Thus the sphere becomes invisible resulting in an almost zero cross section.}
\label{fig:3}
\end{figure}

\begin{figure}[ht]
\centering
\includegraphics[width = 0.60\textwidth]{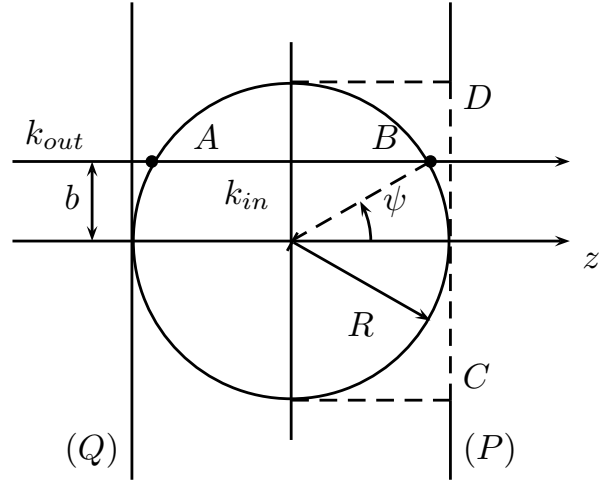}
\caption{Definition of the variables used in the text to calculate the extinction efficiency factor of light over a spherical particle of radius $R$.}
\label{fig:4}
\end{figure}

The idea to model the phenomenon is to consider the incident particle as a plane wave with one part going through the target and another which is not (Figure~\ref{fig:3}). The two parts recombine behind the target and then interfere with each other, producing the oscillating behaviour. Depending upon the impact parameter $b$, light rays going through the drop accumulate a phase shift (Figure~\ref{fig:4}). The calculation of $Q_{\rm e,R}$ is given in the~\ref{app:ramsauer},
\begin{equation}
\label{eq:1}
Q_{\rm e, R} =2 \left (1+\frac{n-1}{\Delta_{\rm R}} \right)^2 \left[1 - 2\,\frac{\sin\Delta_{\rm R}}{\Delta_{\rm R}}+\left(\frac{\sin\Delta_{\rm R}/2}{\Delta_{\rm R}/2}\right)^2\right],
\end{equation}
where $\Delta_{\rm R} = 2R\,(k_{\rm in}-k_{\rm out}) = 2R\,(n-1)\,k_{\rm out} = 2x\,(n-1)$. Note that the first multiplicative factor of Equation~\ref{eq:1} explains the fact that the amplitude of the large oscillations grows with $n$. The extension efficiency factors, when plotted against $\Delta_{\rm R}$, show a universal shape (see Figure~\ref{fig:1}.(b)). Our model and the Mie prediction are in good agreement for the three refractive indices used in Figure~\ref{fig:1}.

A small fraction of light, internally reflected (IR) twice, will also contribute to the forward flux. The amplitude of the internal reflection coefficient $r =(n-1)/(n+1)$ is small ($r=1/7$ for visible light upon water drop in air). The phase shift will be roughly $\Delta_{\rm IR} = 2 \times 2\,R \times k_{\rm in} = 4R\,n\,k_{\rm out} = 4\,n\,x$. Thus the main Ramsauer extinction efficiency factor $Q_{\rm e,R}$ is modulated by the internal reflection factor
\begin{equation}
\label{eq:2}
F_{\rm IR} = 1+r^2 \cos \Delta_{\rm IR} = 1 + r^2 \cos (4\,n\,x).
\end{equation}

Finally, the global extinction efficiency factor is
\begin{equation}
\label{eq:3}
Q_{\rm e} = Q_{\rm e,R} \, \times \, F_{\rm IR}.
\end{equation}

\section{Application of the Ramsauer approach to the scattering of light over a drop of water and comparison with the Mie solution}
\label{sec:application}
\begin{figure}[ht]
\centering
\includegraphics[width = 0.75\textwidth]{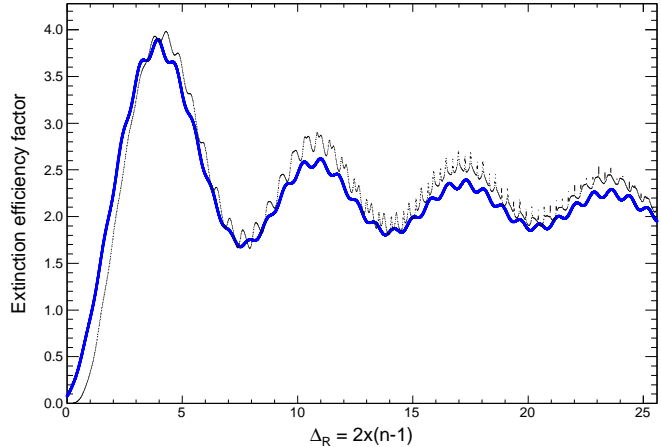}
\caption{Comparison of the extinction efficiencies, the Mie prediction (thin line) (calculated according~\cite{Gliwa}), and the prediction of the Ramsauer model and multiple internal reflections (thick line). $x$ is given by the relation $x = 2 \pi R/ \lambda = k_{\rm out}R$.}
\label{fig:5}
\end{figure}

If we consider visible light ray ($\lambda = 0.6 \, \mu$m) on a drop of water ($n = 1.33$) in air, the Mie and the Ramsauer predictions are compared in Figure~\ref{fig:5}. The Mie solution is a series with more than $600$ terms that has to be computed for each value of the abscissa. On the contrary, the Ramsauer approach is a purely analytical function (Equations~(\ref{eq:1}),~(\ref{eq:2}),~(\ref{eq:3})) with clear physical concepts. The Ramsauer model reproduces quite well both the amplitude and the peak positions of the Mie prediction, except for the main peak position which is lower by almost $10\%$. At small parameter values, the two curves differ. The analytical formula~(\ref{eq:1}), obtained under the light ray approximation, is not expected to be a good description of the reality when the wavelength of the incident light is very much larger than the size of the droplet. This is the Rayleigh regime where the cross section behaves as $1/ \lambda^{4}$. In terms of the variable $x$, this implies, as $x \ll 1$, a $x^4$ behaviour whereas our formula approaches a parabola. Even though the small ripples are also present, they look more attenuated in the Ramsauer curve. Another particularity of our model is that it justifies the ratio between the Ramsauer-pseudo period and internal reflection-pseudo period. Since one has already determined the phase difference for each case, it is straightforward to derive their ratio using  the equations that require constructive interferences for the $k_{\rm out}\,R$-axis,
\begin{equation}
\label{eq:4}
\frac{\Delta_{\rm R}}{\Delta_{\rm IR}} = \frac{n-1}{2\,n}.
\end{equation}
In the case of a raindrop with $n=4/3$, there is thus a factor $8$ between the two pseudo periods. Therefore, we have a simple physical explanation for the origin of the oscillations at two frequencies, and we have derived the ratio of their periods.

The Ramsauer framework allows also, through the Huygens-Fresnel principle, to calculate the unpolarized differential cross section. This is developped in~\ref{app:differentialcrosssec} where Ramsauer and Mie are compared on a particular example.

\section{Conclusion}
We have shown that within the experimental errors the Ramsauer effect predicts an extinction efficiency factor comparable to the one given by the Mie solution. The large oscillatory behaviour is understood as the consequence of the interference between the fraction of light going through and the fraction avoiding the drop. The origin of the small ripples can be traced back to the internally reflected light interfering with the other two components.

\ack
The authors thank their collaborators B.~K\'{e}gl, P.~Eschstruth, D.~Veberi\v{c} and the referees for the improvements due to their comments on the manuscript.

\appendix
\setlength{\mathindent}{0pt}

\section{The Mie solution}
\label{app:mie}
The theory describing the scattering of an electromagnetic plane wave by a homogeneous sphere was originally presented by Gustav Mie~\cite{Mie}. Particles with a size comparable to the wavelength of visible light are relatively common in nature. The laws describing the total scattered intensity as a function of the incident wavelength and the characteristics of the particle are much more complex than the one for the Rayleigh scattering. The $1/\lambda ^4$ dependence of the total scattered intensity in the Rayleigh case is not true anymore in the general case for the Mie solution. Thus the extinction efficiency factor  $Q_{\rm e} = \sigma_{\rm tot}/(\pi R^2) $ depends on the radius and the relative refractive index of the particle.

The Mie theory uses Maxwell's equations to obtain a wave propagation equation for the electromagnetic radiation in a three dimensional space, with appropriate boundary conditions at the surface of the sphere. The extinction efficiency factor obtained is
\begin{equation}
\label{eq:A1}
Q_{\rm e} = \frac{2}{x^2} \sum_{\ell=1}^{\infty} (2 \ell+1) \mathop{\mathrm{Re}} (a_{\ell}+b_{\ell}).
\end{equation}

The Mie scattering coefficients $a_{\ell}$ and $b_{\ell}$ are functions of the size parameter $x = 2 \pi R/ \lambda = k_{\rm out}R$ and of the relative index of refraction $n$,
\begin{eqnarray}
\label{eq:A2_3}
a_{\ell} = \frac{x \psi _{\ell}(x) \psi^{'}_{\ell}(y) - y \psi^{'}_{\ell}(x) \psi _{\ell}(y)}{x \zeta _{\ell}(x) \zeta ^{'}_{\ell}(y) - y \zeta ^{'}_{\ell}(x) \zeta _{\ell}(y)}, \\
b_{\ell} = \frac{y \psi _{\ell}(x) \psi^{'}_{\ell}(y) - x \psi^{'}_{\ell}(x) \psi _{\ell}(y)}{y \zeta _{\ell}(x) \zeta ^{'}_{\ell}(y) - x \zeta ^{'}_{\ell}(x) \zeta _{\ell}(y)},
\end{eqnarray}
where $y=n\,x$. $\psi _{\ell}(z)$, $\zeta _{\ell}(z)$ are the Riccati-Bessel functions (the prime denotes differentiation with respect to the argument) related to the spherical Bessel functions $j_{\ell}(z)$ and $y_{\ell}(z)$ through the equations
\begin{eqnarray}
\label{eq:A4_5}
\psi _{\ell}(z) = z  j_{\ell}(z), \\
\zeta _{\ell}(z) = z  j_{\ell}(z) - \rmi z y_{\ell}(z).
\end{eqnarray}

Numerically, an infinite sum cannot be computed. Thus it is necessary to truncate the series and keep enough terms to obtain a sufficiently accurate approximation. The criterion developed by Bohren CF in~\cite{Bohren} was obtained by extensive computations. The number of required terms $N$ has to be at least the closest integer to $x + 4x^{1/3} + 2$. For instance, for a raindrop of $50\, \mu$m radius and a visible wavelength of $0.6\, \mu$m, the number of required terms is $N=558$. Nowadays, the computers have reached a point where the computing time is no longer a problem, nevertheless it is interesting to compare to an approximate closed form method in order to obtain some insights of the different physical processes. The plots of the Mie solutions are obtained using the \textit{Fast Mie Algorithm} of Pawel Gliwa~\cite{Gliwa} for total cross sections and the \textit{MiePlot} program of Philip Laven~\cite{Laven} for differential cross sections.

\section{Determination of the extinction efficiency factor: the optical theorem}
\label{app:ramsauer}
In the scattering theory, the wave function far away from the scattering region must have the form
\begin{equation}
\label{eq:B1}
\Psi (\vec{r}) =  \rme ^{\rmi \vec{k} \vec{r}} + f(\theta) \frac{\rme ^{\rmi k r}}{r}.
\end{equation}

The optical theorem reads as
\begin{equation}
\label{eq:B2}
\sigma _{\rm tot}=\frac{4\pi}{k_{\rm out}}\mathop{\mathrm{Im}}f(\theta = 0),
\end{equation}
where $f(\theta = 0)$ is the forward scattering amplitude. Under the approximation for the scattering of a scalar (spinless) wave on a spherical and symmetric potential, the scattering amplitude at a given polar angle $\theta$ can be written as a sum over partial waves amplitudes, each of different angular momentum $\ell$ as follow
\begin{equation}
\label{eq:B3}
f(\theta)=\frac{1}{2 \rmi k_{\rm out}}\sum_{\ell =0}^\infty(2 \ell +1)P_{\ell} (\cos\theta)\left[\eta_{\ell}\, \rme ^{2 \rmi \delta_{\ell}}-1\right],
\end{equation}
where $\eta_{\ell}$ is the inelasticity factor ($\eta_{\ell} = 1$ in our case of a non absorbing sphere), $\delta_{\ell}$ is the phase shift ($\delta_l$ is real for a pure elastic scattering), and $P_{\ell}(\cos\theta)$ is the Legendre polynomial.

From Figure~\ref{fig:4}, $\delta_{\ell}=(k_{\rm in}-k_{\rm out})R\cos\psi$. Under the approximation of the forward scattering, $P_\ell(\cos\theta)\rightarrow 1$. Let us introduce the impact parameter $b=R \sin\psi$, such that $\ell=b k_{\rm out}$ (from the Bohr momentum quantization : $b p=\ell \hbar$ and $p=\hbar k$). By substituting the discrete sum over $\ell$ by a continuous sum over $\ell$ or over the impact parameters 
$$
\sum_{\ell} \rightarrow \int \rmd \ell \rightarrow k_{\rm out}\int \rmd b ,
$$
the forward scattering amplitude can be written in term of the impact parameter

\begin{equation}
\label{eq:B4}
f(\theta = 0)=\frac{k_{\rm out}}{\rmi} \int_{0}^R \left( \rme ^{\rmi 2(k_{\rm in}-k_{\rm out})R\cos \psi}-1\right) b \rmd b ,
\end{equation} 
or in term of the angle $\psi$
\begin{equation}
\label{eq:B5}
f(\theta = 0)= \frac{k_{\rm out} R^2}{2\rmi} \int_{0}^{1} \left( 1- \rme ^{\rmi 2(k_{\rm in}-k_{\rm out})R\cos \psi}\right) \rmd \cos^2\psi .
\end{equation}

Then the expression for the total cross section is obtained from Equation~(\ref{eq:B4}) and setting $w=\cos\psi$, 

\begin{eqnarray}
\eqalign{
\sigma_{\rm tot}&=\mathop{\mathrm{Im}}\left[ \rmi \,4\pi R^2\int_0^{1}w\left(1 - \rme ^{ \rmi (k_{\rm in}-k_{\rm out})2Rw}\right) \rmd w\right] \\
&=\mathop{\mathrm{Im}}\left[ \rmi \,2\pi R^2- \rmi \, 4\pi R^2\int_0^1 w\, \rme ^{ \rmi (k_{\rm in}-k_{\rm out})2Rw} \rmd w\right].}
\label{eq:B6}
\end{eqnarray}

Then, from integration by parts, we obtain
\begin{eqnarray}
\eqalign{
\fl \sigma_{\rm tot}&=\mathop{\mathrm{Im}}\left[ \rmi \,2\pi R^2- \rmi \, 4\pi R^2\left(\left[w\frac{ \rme ^{ \rmi (k_{\rm in}-k_{\rm out})2Rw}}{ \rmi (k_{\rm in}-k_{\rm out})\,2R}\right]_0^1-\int_0^1\frac{ \rme ^{ \rmi (k_{\rm in}-k_{\rm out})2Rw}}{ \rmi (k_{\rm in}-k_{\rm out})\,2R} \rmd w\right)\right]\\
&=\mathop{\mathrm{Im}}\left[ \rmi \,2\pi R^2- \rmi \, 4\pi R^2\left(\frac{ \rme ^{ \rmi (k_{\rm in}-k_{\rm out})2R}}{ \rmi (k_{\rm in}-k_{\rm out})\,2R}+\frac{ \rme ^{ \rmi (k_{\rm in}-k_{\rm out})2R}-1}{(2R)^2\,(k_{\rm in}-k_{\rm out})^2}\right)\right]\\
&=2\pi R^2-4\pi R^2\left[\frac{\sin\left(2R\,(k_{\rm in}-k_{\rm out})\right)}{2R\,(k_{\rm in}-k_{\rm out})}-\frac{1}{2}
\left[ \frac{\sin\left(R\,(k_{\rm in}-k_{\rm out})\right)}{R\,(k_{\rm in}-k_{\rm out})}\right]^2
\right].
}
\label{eq:B7}
\end{eqnarray}

If we introduce the parameter $\Delta_{\rm R} = 2R\,(k_{\rm in}-k_{\rm out})$, the analytic expression for the cross section becomes
\begin{equation}
\sigma_{\rm tot} = 2\pi R^2\left[1 - 2\,\frac{\sin\Delta_{\rm R}}{\Delta_{\rm R}}+\left(\frac{\sin\Delta_{\rm R}/2}{\Delta_{\rm R}/2}\right)^2\right].
\label{eq:B8}
\end{equation}

In this expression, the total cross section approaches $2\pi R^2$ at high energies. It is the so called "black disk" approximation. Nevertheless, from a purely geometrical viewpoint, a collision occurs if the distance between the two sphere centers is less than $r_{1}+r_{2}$, where $r_{1}$ and $r_{2}$ are the radii of the two spheres. So the geometrical cross section is equal to $\pi (r_{1}+r_{2})^2$. In our case, the photon may be considered as a particle with a diameter equal to its reduced wavelength $\bar{\lambda} = \lambda / 2 \pi$. Consequently, Equation~(\ref{eq:B8}) becomes
\begin{equation}
\sigma_{\rm tot} = 2\pi \left(R+\frac{\lambda}{4\pi} \right)^2\left[1 - 2\,\frac{\sin\Delta_{\rm R}}{\Delta_{\rm R}}+\left(\frac{\sin\Delta_{\rm R}/2}{\Delta_{\rm R}/2}\right)^2\right].
\label{eq:B9}
\end{equation}

Finally, an expression for $\Delta_{\rm R}$ as a function of the index of refraction $n$ is needed. According to Maxwell approach, it is $k_{\rm in} = n\,k_{\rm out}$. With this relation and the fact that $\lambda = 2 \pi/k_{\rm out}$
\begin{eqnarray}
\eqalign{
\sigma_{\rm tot} = 2\pi R^2 \left (1+\frac{n-1}{\Delta_{\rm R}} \right)^2 \left[1 - 2\,\frac{\sin\Delta_{\rm R}}{\Delta_{\rm R}}+\left(\frac{\sin\Delta_{\rm R}/2}{\Delta_{\rm R}/2}\right)^2\right] \\
{\rm or} \\
Q_{\rm e, R} =2 \left (1+\frac{n-1}{\Delta_{\rm R}} \right)^2 \left[1 - 2\,\frac{\sin\Delta_{\rm R}}{\Delta_{\rm R}}+\left(\frac{\sin\Delta_{\rm R}/2}{\Delta_{\rm R}/2}\right)^2\right].}
\label{eq:B10}
\end{eqnarray}

This result is derived under the approximation of a scalar light whereas the Mie solution has to do with the full vectorial light. Therefore our formula applies only to unpolarized light scattering.

\section{Ramsauer solution for the differential cross section}
\label{app:differentialcrosssec}
Let us first have a reminder about the Huygens-Fresnel principle. The propagation of light can be described with the help of virtual secondary sources. Every point of a chosen wavefront becomes such a secondary source and the light amplitude at any chosen location is the integral of all these sources emitting spherical waves towards that observation point. The secondary sources are driven by the incident light. Their amplitude is 
$$
\frac{1}{\rmi \lambda} \frac{1+\cos\theta}{2} A_{\rm incident} \rmd S
$$
where $\lambda$ is the wavelength and $A_{\rm incident}$ is the amplitude of the incoming light. $\theta$ is the angle between the direction of the incident light and the direction from the virtual source to the observation point. At last the elementary area where the secondary source stands is $\rmd S$.

We decide to use the plane (P), shown in Figure~\ref{fig:4}, as the location of our secondary sources. The shadow of the water sphere on the plane (P) is the disk CD. If the plane (Q) is chosen as the origin of the phases, the incident light is a plane wave which, on the plane (P), has the value $\rme^{\rmi 2 k R}$. The amplitude distribution of the secondary sources can be split into two parts: an undisturbed plane wave and a perturbation over CD only. The undisturbed plane wave, when integrated, gives a Dirac delta function in the forward direction. Thus the differential cross section comes from the sources on CD. The observation points are very far away so that to get the amplitude at an angle $\theta$ we just sum all directions parallel to $\theta$.

Let $\rho$ and $\phi$ be the polar coordinates in the plane (P). The amplitudes of the virtual sources in the disk CD are then
\begin{eqnarray}
\eqalign{
A_{\rm disk}(\rho)&= \left[-\rme^{\rmi 2k R} + \rme^{\rmi n 2 k \sqrt{R^2-\rho^2}}\, \rme^{\rmi k (2R-2\sqrt{R^2-\rho^2})} \right]\frac{-\rmi}{\lambda} \frac{1+\cos\theta}{2}\\
&=\rmi \frac{k}{2 \pi}\rme^{\rmi 2 k R}\left[1- \rme^{\rmi 2 k \sqrt{R^2-\rho^2}\;(n-1)}\right] \frac{1+\cos\theta}{2} .
}
\label{eq:C1}
\end{eqnarray}

The resulting scattering amplitude of Equation~(\ref{eq:B1}), now refered as $f_{\rm sphere}(\theta)$, is obtained after moving the origin of the phases to the plane (P). This is simply done through a multiplication by $\rme^{-\rmi 2 k R}$
\begin{eqnarray}
\eqalign{
f_{\rm sphere}(\theta)&=\rme^{-\rmi 2 k R} \int_0^{2\pi} \rmd \phi \int_0^R A_{\rm disk}(\rho) \, \rho \, \rme^{\rmi k \rho \cos\phi\sin\theta} \rmd \rho \\
&= \rmi \frac{k}{2 \pi}\frac{1+\cos\theta}{2} \int_0^{2\pi} \int_0^R \rme^{\rmi k \rho \cos\phi\sin\theta} \left[1- \rme^{\rmi 2 k \sqrt{R^2-\rho^2}\;(n-1)}\right] \rmd \rho \rho \rmd \phi ,
}
\label{eq:C2}
\end{eqnarray}
where the term $\rho \cos\phi\sin\theta$ represents the path length difference between a source in the disk and a source at disk center for an observer placed at an infinite distance from the disk.

But the integral over the angle $\phi$ is a Bessel function
\begin{equation}
J_0(u) = \frac{1}{2 \pi} \int_0^{2\pi} \rme^{\rmi u \cos \phi} \rmd \phi .
\label{eq:C3}
\end{equation}

Thus Equation~(\ref{eq:C2}) becomes
\begin{equation}
f_{\rm sphere}(\theta)= \rmi k\frac{1+\cos\theta}{2} \int_0^R \rho J_0(k \rho \sin \theta) \left[1- \rme^{\rmi 2 k \sqrt{R^2-\rho^2}\;(n-1)}   \right] \rmd \rho .
\label{eq:C4}
\end{equation}
This is as far as we can go with usual functions. Note that the scattering function in the forward direction is
$$
f_{\rm sphere}(\theta = 0) = \rmi k \int_0^R \left[1- \rme^{\rmi 2 k \sqrt{R^2-\rho^2}\;(n-1)} \right] \rho \rmd \rho ,
$$
which is exactly Equation~(\ref{eq:B4}).

We can get a closed form formula if the sphere is replaced by a disk of radius $R$ and of height $2R$ along $z$. Equation~(\ref{eq:C4}) simplifies into
\begin{eqnarray}
\eqalign{
f_{\rm disk}(\theta)&=\rmi k \frac{1+\cos\theta}{2} \int_0^R \rho J_0(k \rho \sin \theta) \left[1- \rme^{\rmi 2 k R\,(n-1)}\right] \rmd \rho  \\
&= \rmi k \frac{1+\cos\theta}{2} \left[1- \rme^{\rmi 2 (n-1)k R}\right]  \int_0^R \rho J_0(k \rho \sin \theta) \rmd \rho \\
&= \rmi \frac{1+\cos\theta}{2} \rme^{\rmi \frac{2 (n-1)\,k R}{2}} (-2 \rmi)\sin \left[\frac{2 (n-1)\,kR}{2} \right] \frac{R}{\sin \theta} J_1(k R \sin \theta),
}
\label{eq:C5}
\end{eqnarray}
since the integral can be simplified by the fact that 
$$
\int_0^R u J_0(u) \rmd u = R \, J_1(R).
$$

Then the differential cross section reads
\begin{equation}
\frac{\rmd \sigma}{\rmd \theta}=\left| f_{\rm disk}(\theta)\right|^2 = R^2\left[kR \frac{1+\cos \theta}{2} \sin\left[(n-1)kR \right]\frac{2 J_1(kR\sin \theta)}{kR \sin \theta} \right]^2.
\label{eq:C6}
\end{equation}

In order to compare predictions from Ramsauer approach and Mie, Equation~(\ref{eq:C6}) can be rewritten as a function of the size parameter $x=k R$ and normalized by $\pi R^2$
\begin{equation}
\frac{1}{\pi R^2} \frac{\rmd \sigma}{\rmd \theta}=\frac{1}{\pi} \left[x \frac{1+\cos \theta}{2} \sin\left[x(n-1)\right]\frac{2 J_1(x\sin \theta)}{x\sin \theta} \right]^2 .
\label{eq:C7}
\end{equation}

When the relative index of refraction $n$ is equal to $1$, the differential cross section is null which is normal. Also, if the phase shift in the sphere is not far from a multiple of $2 \pi$, then again the cross section is null. This is the Ramsauer effect again: when the outside and the inside are in phase, they are producing an almost invisible sphere.

\begin{figure}[ht]
\centering
\includegraphics[width = 0.75\textwidth]{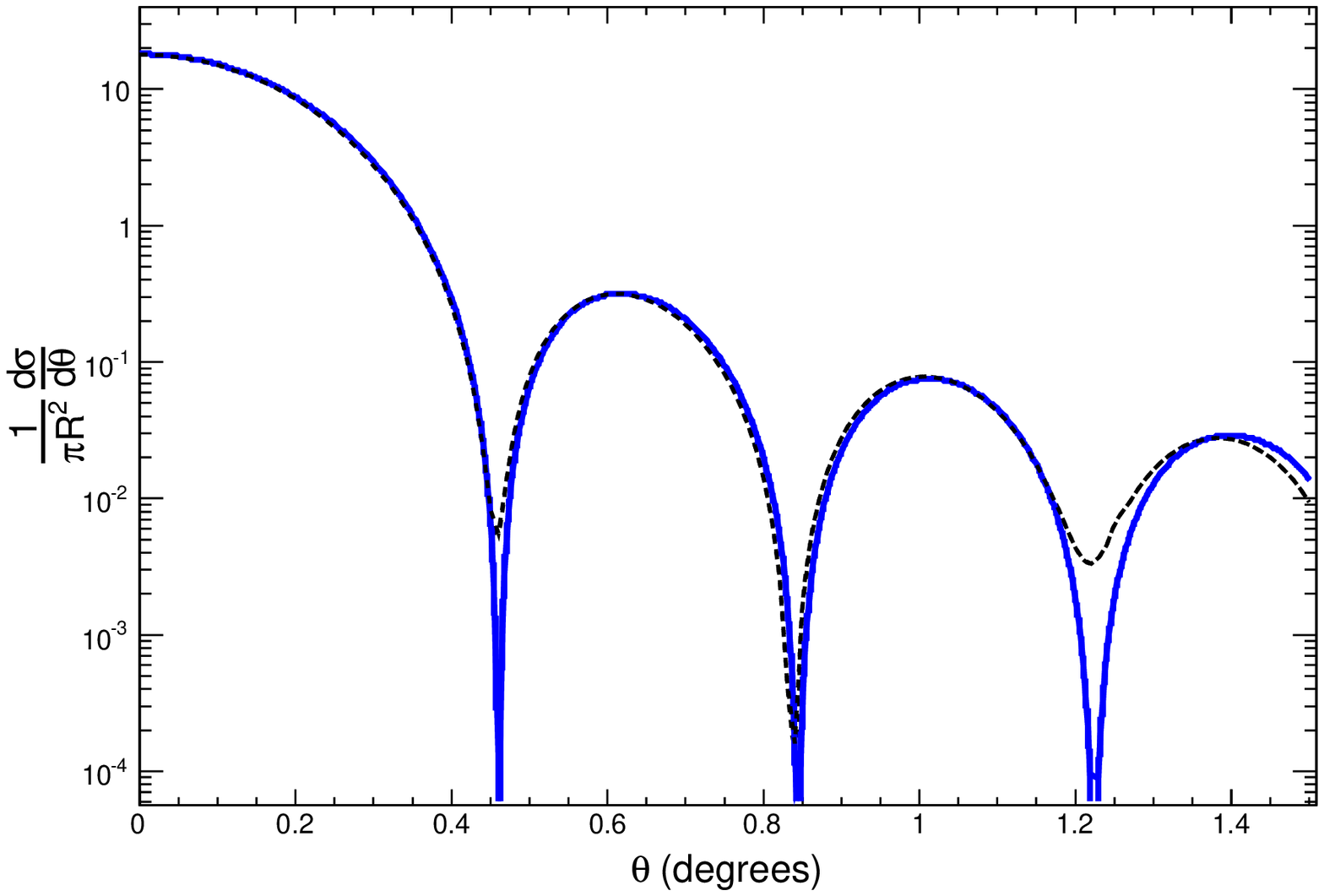}
\caption{Angular scattering diagram for $45.45$~$\mu$m droplet ($n=1.33$) as obtained by Ramsauer (thick line) and Mie (dashed line) (calculated according~\cite{Laven}) approaches for angles between $0^{\rm o}$ and $1.5^{\rm o}$. The wavelength taken is equal to $0.6$~$\mu$m.}
\label{fig:6}
\end{figure}

Figure~\ref{fig:6} shows the differential cross sections from Mie and from Ramsauer for a drop of water of $45.45~\mu$m radius, in air, at an incident wavelength of $0.6~\mu$m. As in the case of the total cross section, we find that the Ramsauer formula for a disk is a fair approximation of Mie for a sphere.

\setlength{\mathindent}{70pt}

\end{document}